# MFCPji & MFIDji: New ImageJ Macros To Analyze Structure Formed In Magnetic Nanofluid

Urveshkumar Soni, Rucha P Desai[a]

P. D. Patel Institute of Applied Sciences, Charotar University of Science and Technology (CHARUSAT), CHARUSAT Campus, Changa, 388421, Gujarat, India.

[a]ruchadesai.neno@charusat.ac.in

**ABSTRACT**

The aqueous magnetic nanofluid consists of superparamagnetic nanoparticles, with a typical size 10-12 nm. On the application of the magnetic field, these nanoparticles align heterogeneously and form a chain or chain-like structure. This structure is observed using a microscope. Although such chain or microstructure formation is well reported in many articles, the method to identify and determine chain parameters, e.g., chain length, width, and associated counts, are scare. Similarly, inter-chain or successive distance is one of the critical parameters for the development of magnetic nanofluid based devices. This paper describes Magnetic Field induced Chain Parameters (MFCP) and Magnetic Field induced Interchain Distance (MFID) a set of new ImageJ methods to identify and determine (i) chain length, width, and associated counts, along with (ii) successive distance of the chains respectively in the magnetic nanofluid. This utilises a macro files such as MFCPji.txt and MFIDji.txt for ImageJ, which can be used on microscopic image of magnetic nanofluid without and with application of magnetic field. The method requires no specialised scientific equipment and can be run entirely using free to download software. The examples of microstructure formations in two different magnetic fluids (A & B) are discussed. The results of the associated weighted average chain width and counts, as well as the successive distance between the chains, are reported. The chain parameters are useful to determine diffraction grating angle. The MFCPji and MFIDji macros has been integrated into a macro tool set that can be configured to be run on ImageJ start up. The MFCPji and MFIDji are available from the following Uniform Resource Locator (URLs):

https://github.com/urveshsoni/ImageJ---Macros

https://ruchadesailab.wordpress.com/publication/

**INTRODUCTION**

The magnetic nanofluid or ferrofluid is a stable colloidal suspension of superparamagnetic particles (~ 10*nm*) suspended in a polar/non-polar carrier such as water, kerosene, oil [1]. The superparamagnetic nanoparticles (NPs) stay at a random orientation at zero magnetic field. During the synthesis of the polar based magnetic fluid, surfactant coating can be done in two ways: either the surfactant directly coats on the aggregates of NPs or surfactant coats on the monomer/dimer and eventually form aggregates. Therefore, collective aggregate behavior is dominated in the aqueous magnetic fluid [2]. On the other side, non-polar based fluid has less tendency to form aggregates during the synthesis, and hence individual particle behavior is dominated.

The water-based magnetic fluid exhibits nearly isotropic structure (including pre-existed aggregates) at a zero magnetic field. The presence of pre-existed aggregates can be detected by using appropriate tools (for example, transmission electron microscope (TEM), scanning electron microscope (SEM), up-right/inverted/bring-field microscope) [2–6]. Under the magnetic field, individual NPs/aggregates of magnetic fluid start aligning in the magnetic field direction, form thin/thick chains/chain-like/column-like structure, and leading to uniaxial anisotropic structure [2,6–8]. Generally, uniaxial anisotropy reverts to the original isotropic structure on switching off the magnetic field, which is the vital aspect while seeking the potential applications of magnetic fluid. Numerous research groups have determined magntic field induced microstructure parameters theoretically [9–11] and with the help of computer simulations such as molecular dynamics and brownian dynamics [12,13] in the magnetic nanofluid. The magnetic microstructure formation is the base for the functioning of several optical devices such as magnetic fluid switch, grating, modulator, sensor, lens [14]. Thus, it is essential to determine the magnetic field induced microstructures parameters (length, width, successive distance).

The analysis/processing of images is possible through programming language and/or various software such as Matlab [15], Python [16], and ImageJ [17]. Several researchers developed methods and plugin to analyze material[18] and biological samples, i.e., automatic cell counting[19], leaf shape measurement[20], and tissue analysis[21] using ImageJ software. JColloids is used for the identification of the microstructures in the magnetorheological (MR – suspension of micron-sized particles) fluid [22,23]. It recognizes the area (not length-width) of the structure (typical >30 µm$^2$), XY position coordinates. The image imported in the ImageJ environment can be directly processed using JColloids add-on. Perhaps it is designed to identify large structures, and hence it is not able to identify the presence of structure $< 30 \ \mu m^2$. Such a small microstructure is expected in the magnetic fluid. To overcome this limitation, we propose here an alternative methods to determine microstructure parameters such as length, width, and successive distance in the magnetic nanofluid. The motivation for this is based on the method to measure the length and width of an orthogonal object (rice seeds) using ImageJ [24], and method to measure successive cell behavior (biological systems) using ImageJ [25]. However, a method to determine magnetic field induced length, width, and successive distance of the chains in the magnetic fluid system is not reported in the literature. Here, we report MFCPji and MFIDji macros to determine (i) chain length, width, and total counts, and (ii) successive distance of the chains respectively in ImageJ. The results of the proposed method are being compared with the output obtained using JColloids. The obtained parameters would be useful for the fabrication of magnetic fluid based tunable grating and other optical devices.

**EXPERIMENTAL PROCEDURE**

**Magnetic Field Induced Microscopy**

An inverted microscope (IM7200, Meiji Techno) attached to the charged coupled device (CCD) (ProgRes, Jenoptics) is used to capture the images (2080 × 1542 pixels). Initially, an image is captured at a zero magnetic field. Subsequently, images are captured at a constant magnetic field (H = 0.046T) for different exposure time (1 to 5 minutes). The captured images are stored in a Bitmap (*.BMP) format. The bitmap provides better resolution compared to

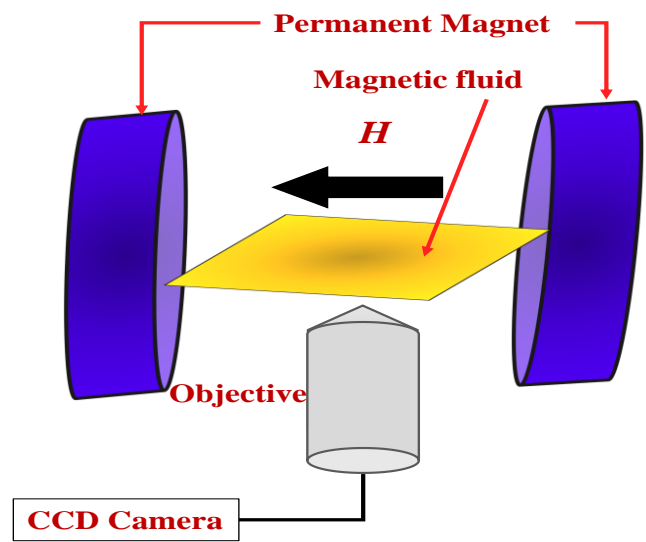

Figure 1: The schematic diagram of the magnetic field induced microscopy at constant magnetic field 0.046 T.

Joint Photographic Expert Group (JPEG), Portable Network Graphics (PNG), and Graphics Interchange Format (GIF). The magnetic field induced microstructure in the water-based magnetic nanofluids A and B are imaged at a constant magnetic field (H) of 0.046 T using 20X objective lens and a CCD camera. Figure 2(a) shows a typical image of fluid B at H = 0 T, revealing the presence of pre-aggregates. Figure 2(b) demonstrates a typical image of the microstructure formation after applying H = 0.046 T for two minutes. The magnetic field applied is perpendicular to the sample, as well as the incident light (figure 1). The software (ProgRes) used for image capturing is pre-calibrated using the standard. The minimum resolution obtained for the 20X objective lens (N.A. 0.4) and a CCD camera was <1µm.

**Work Flow of the MFCPji and MFIDji Macros**

In this work, two automatic image analysis methods such as MFCPji and MFIDji macros were developed. The macros's process, described herein three parts: (a) commons steps, (b) method to determine the magnetic field induced structure parameters (such as length, width, count) (i.e., MFCPji), and (c) method to determine the magnetic field induced successive distance (i.e., MFIDji) in magnetic nanofluid using ImageJ (updated version - Fiji v1.8.0) software[26]. To utilise the MFCPji and MFIDji macros, save the "*.ijm" file to ImageJ's "Plugins" folder. After that, the macro will appear in the "Plugins" menu. When the macro is executed, it automatically processes microscopic images without and with the assistance of a magnetic field, taking only a few seconds to examine the microstructure dimensions. Finally, all results are stored as a ".xls" file suitable for statistical analysis. The process is described below.

**(a)  Common steps**

Figure 2 shows the algorithm of the developed protocol and output of the common steps. When the MFCPji and MFIDji macros are executed, the imageJ software prompt you to insert a microscopic image with no magnetic field (figure 2(a)). After clicking "okay," the software ask to perform calibration. To calibrate the images, insert the scale value using a set scale option (under Analyze menu) and make it global (global will set the scale in all open images). Later on, the ImageJ software prompt to insert a microscopic image under the influence of the magnetic field (figure 2(b)). Figure 2(c) shows the net effect observed by subtracting the zero-field image from the in-field image using a calculator plus (under Process menu) tool. Now, click on the sharp tools (under Process menu) to enhance the intensity of tiny microstructures. Convert the color image into 16-bit greyscale (Image >Type > 16 bit) (figure 2(d)). Apply a threshold tool (Image > Adjust > Auto-threshold) on image to make the microstructures into white (figure 2(e)).

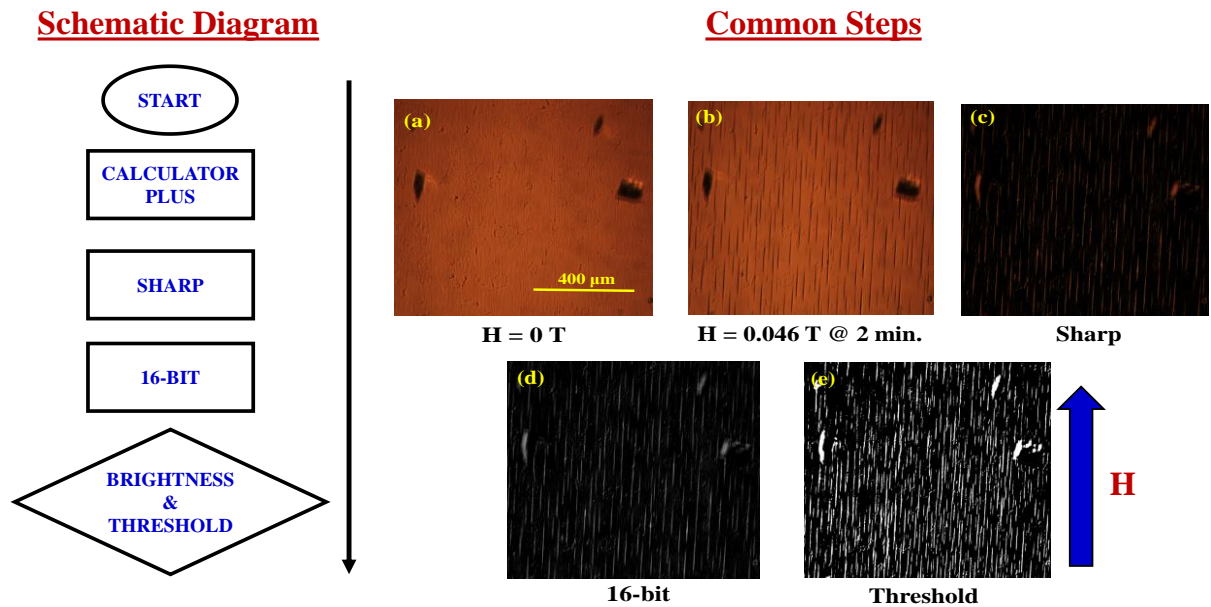

Figure 2: The schematic diagram of the developed algorithm of the common steps and the magnetic fluid at (a) magnetic field (H) = 0T, (b) after applying 0.046 T magnetic field for two minutes, (c) zero-field image subtracted from in-field image (i.e., b-a), and (d) image threshold. The scale bar is 400 μm.

**(b)    MFCPji: Method to determine the magnetic field induced structure parameters**

Figure 3 shows the algorithm of the developed protocol of MFCPji and output of the individual steps. After getting the threshold image, click on the dilate tool (Process > Binary> Dilate) to see the edges of the microstructures clearly (figure 3(f)). Before analyzing the particles, select Analyze > Set Measurements, and select options like area, and bounding rectangles. The option bounding rectangles will not restrict to rectangular or square, but it means a frame or boundary of the structure. Subsequently, select analyze particle tool under the Analyze menu (figure 3(g)). Along with the final image, it will generate two tables, one for summary and other for individual microstructure parameters (such as area, length, width). The data generated can be copied in any of the spreadsheet programs (say Microsoft Excel, Origin, Openoffice spreadsheet).

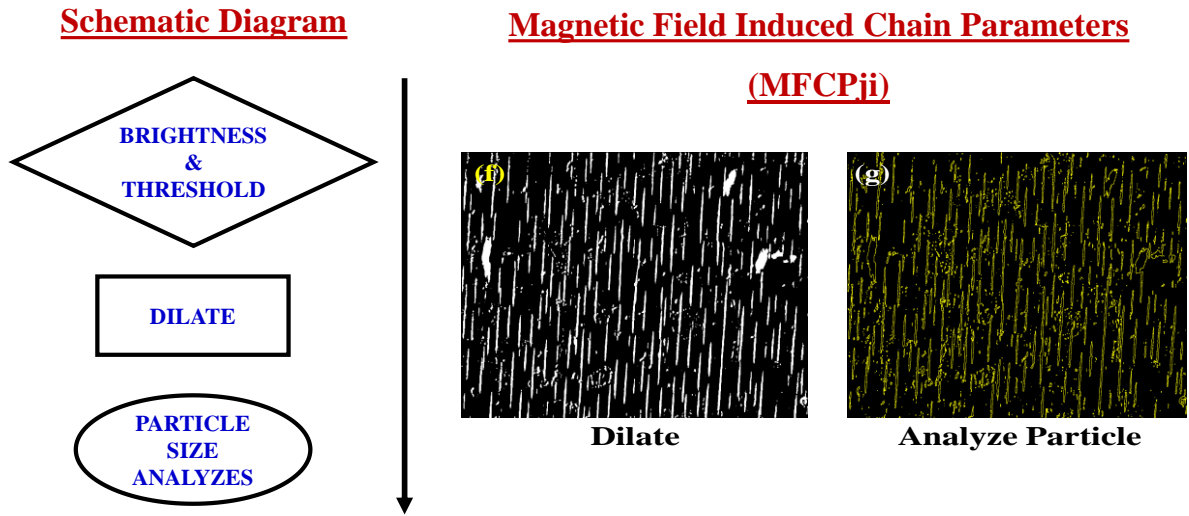

Figure 3: The schematic diagram of the developed algorithm of the magnetic field induced chain parameters and steps to determine magnetic field induced chain parameters (f) dilate, and (g) analyze particle. The scale bar is 400μm.

**(c)  MFIDji: Method to determine the magnetic field induced successive distance**

The algorithm for the developed MFIDji protocol is depicted in Figure 4, along with the output of the sequential steps. Repeat the process until figure 2(e) as mentioned in the common steps. Click on the invert option under Edit menu (figure 4(h)) and apply a distance map (Process > Binary > distance map) (figure 4(i)) tool to measure the successive distance between microstructures. The better outcome can be achieved by converting the image from binary to color using 16_colors plugin (figure 4(j)). Apply the minimum threshold on the image. Now, select Analyze > Set Measurements > Area. Click on Analyze > Measure option (figure 4(k)). It results in a new window with certain parameters like area, mean, minimum, and maximum threshold values. Here the important parameter of our interest is mean, which is the successive distance between microstructures.

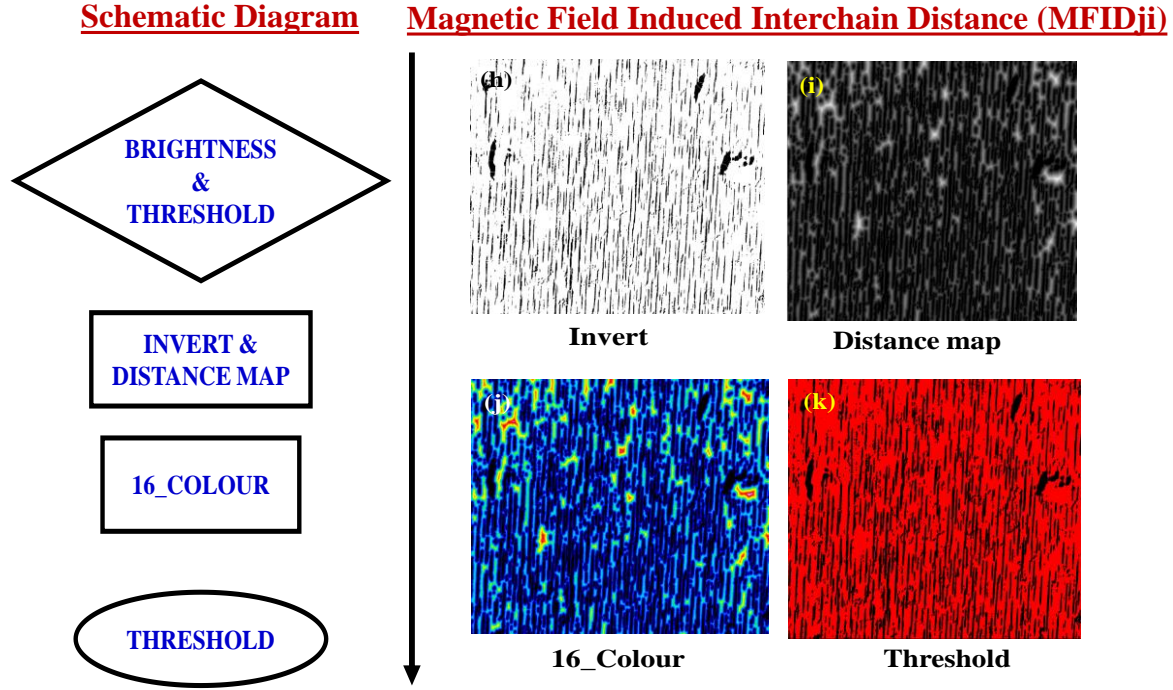

Figure 4: The schematic diagram of the developed algorithm of the magnetic field induced interchain distance and steps to determine the successive distance between microstructures (h) invert, (i) distance map, (j) 16_color, and (k) threshold. The scale bar is 400 μm.

**RESULT AND DISCUSSION**

The aqueous magnetic nanofluid is composed of superparamagnetic iron oxide which separated with the use of a surfactant, specifically lauric acid. The magnetic moment (m) of the superparamagnetic particle is aligned with the direction of the applied magnetic field, forming chains and columns. Halsey and Toor's electrorheological fluid model is more appropriate for understanding magnetic field-induced microstructure formation. When a strong magnetic field is applied, particles in the magnetic nanofluid form chains and columns that reach from one end of the cell to the other. A chain's cross-section will include a single particle, but a column will contain multiple such chains. Without an external magnetic field, Brownian motion drives the particles, and so the net magnetization is zero. The magnetic moment of an individual particle in the presence of an external magnetic field is given by

$$m = \frac{\pi}{6} a^3 \chi_{eff} H$$

where, $a$ is the magnetic nanoparticle diameter, and $\chi_{eff}$ the effective susceptibility. The anisotropic dipolar potential energy of pairs of particles is expressed as, $U_{ij}(r_{ij}, \theta_{ij}) = \frac{m^2 \mu_0}{4\pi} \left( \frac{1 - 3\cos^2\theta_{ij}}{r_{ij}^3} \right)$ with $r_{ij}$ center to center distance between the i[th] and j[th] particles, and $\theta_{ij}$ the angle between the vector $r_{ij}$ and the magnetic field applied. Using this, the relative strength of dipolar interaction in terms of thermal energy is expressed by coupling parameter $\lambda = -\frac{U(a,O)}{k_B T} = \frac{\pi \mu_0 a^3 \chi^2 H_0^2}{72 k_B T}$. The magnetic particles assemble into the aligned structure ($\lambda \gg 1$) on application of the field, forming dipolar chains and exhibits strong Landu-Peierls fluctuation. Halsey and Toor (HT) illustrates the long-range coupling among dipolar chains due to chain formations, and possess an attractive interaction through power-law decay. The HT model was modified by Martin *et al.*,

$$U \sim \frac{m}{a} \langle H^2 \rangle^{1/2} \sim \frac{\chi H (\mu_0 K_B T)^{1/2} a^{5/2}}{\rho^2}$$

Here, $\rho$ is the distance among two chains or columns. This energy can be either attractive or repulsive. The interaction energy per unit length increases with increasing the field and/or decrease in the inter-chain distance resulting in lateral coalesce of two chains. Consequently, the separation distance (ρ) increases and lowering U, and resulting into reduction of overall energy of the system.

Here, two different types of aqueous magnetic fluid is used to demonstrate the magnetic field induced microstructure formation. Time evolution microscopic images of A and B fluids are captured at H = 0.046 T for the initial 5 minutes with 1-minute intervals and processed as described in Figure 1(a-d). Figure 6 shows the threshold images of magnetic fluids A and B as a function of time. It reveals that at an initial one minute, the chains (microstructures) observed in fluid A are long and relatively distributed, while in fluid B structure is small and dense. As expected, the chain length and inter-chain distance increase with the increase of magnetic field exposure time [2]. With the exposure of the magnetic field continuously, the chains coalesce and zip, resulting in long microstructure formation. The eye-view mimics as a continuous structure, but an in-depth analysis of the image will give more explicit information. The methods described above are followed to understand and determine microstructure parameters for both the fluids A & B (images not shown here for brevity). Finally, generated data are processed (given below), and the weighted average length and width of the chain are determined. Following a method to determine chain length and width, the dataset is generated containing area, height, and width. As image analysis works on the concept of pixel identification, data corresponds to minimum length and width (both) obtained here is 1.26 µm. However, such small-sized chains are not prominently visible in the image, contrary to large-sized. The magnetic fluid comprises of heterogeneous distribution of small and large chains. Later contribute mainly to the structure formation, whereas the former reduces the light transmission.

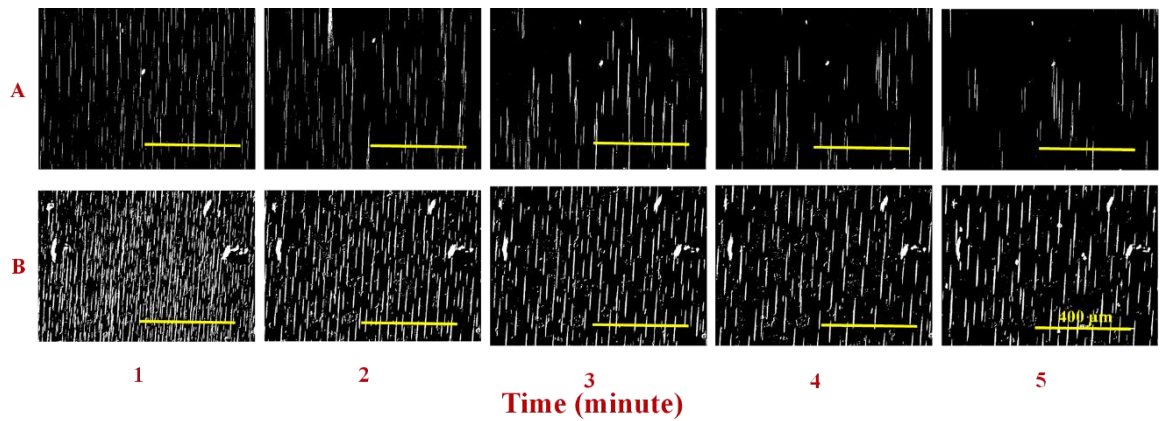

Figure 6: Microscopic confirmation of chain formation after applying image processing for sample A & B. The scale bar is 400 μm.

The fact that small chains orientation either requires a high field and less time or moderate field with long exposure time. Also, the total number of such small chains always remains high. Now in the calculation, if one considers entire distribution, then the weighted average results into false results. It is due to the presence of a large number of small chains, which eventually not contributing to the chain formation effectively within measurement time. Let us consider the following example. In one of the samples, the smallest length obtained is 1.26 µm, while the largest is 280 µm; the rest of the chains are distributed between this range. The weighted average chain length calculated is 17.4 µm. Now, if one considers chains ≥ 10 µm, then the weighted average chain length is 44.6 µm. While comparing these results, the 44.6 µm matches to that of manual observation. Hence, the following analysis is carried out considering chain length ≥ 10 µm.

Figure 7 explains the time-dependent weighted average chain (a) length, (b) width, and (c) number of counts having a length ≥ 10 µm, with a 5% error in all the analysis in the magnetic fluids A and B at H = 0.046 T. In the magnetic fluid A & B, weighted average chain length and width increase with the increasing time with the decrement in the associated counts. In the magnetic fluid A, the chain length ranging from 52 µm to 77 µm, while chain width ranging from 7 µm to 9.5 µm. It indicates that the chains observed in the magnetic fluid A are long and thick. On the contrary, chain length observed in the magnetic fluid B ranging from 45 µm to 65 µm while chain width ranging from 1.8 µm to 3.8 µm. It signifies that the chain observed are long and thin. The chain length obtained using the proposed method agrees with the reported value [7] The longitudinal interaction of chains leads to zipping, and hence chain length increases with time. The transverse interaction increases chain width. The increase in the chain length and width is reflected in the reduction of associated counts. The fact that forces associated with the chain-chain interaction and repulsion results in re-distribution of chain formation.

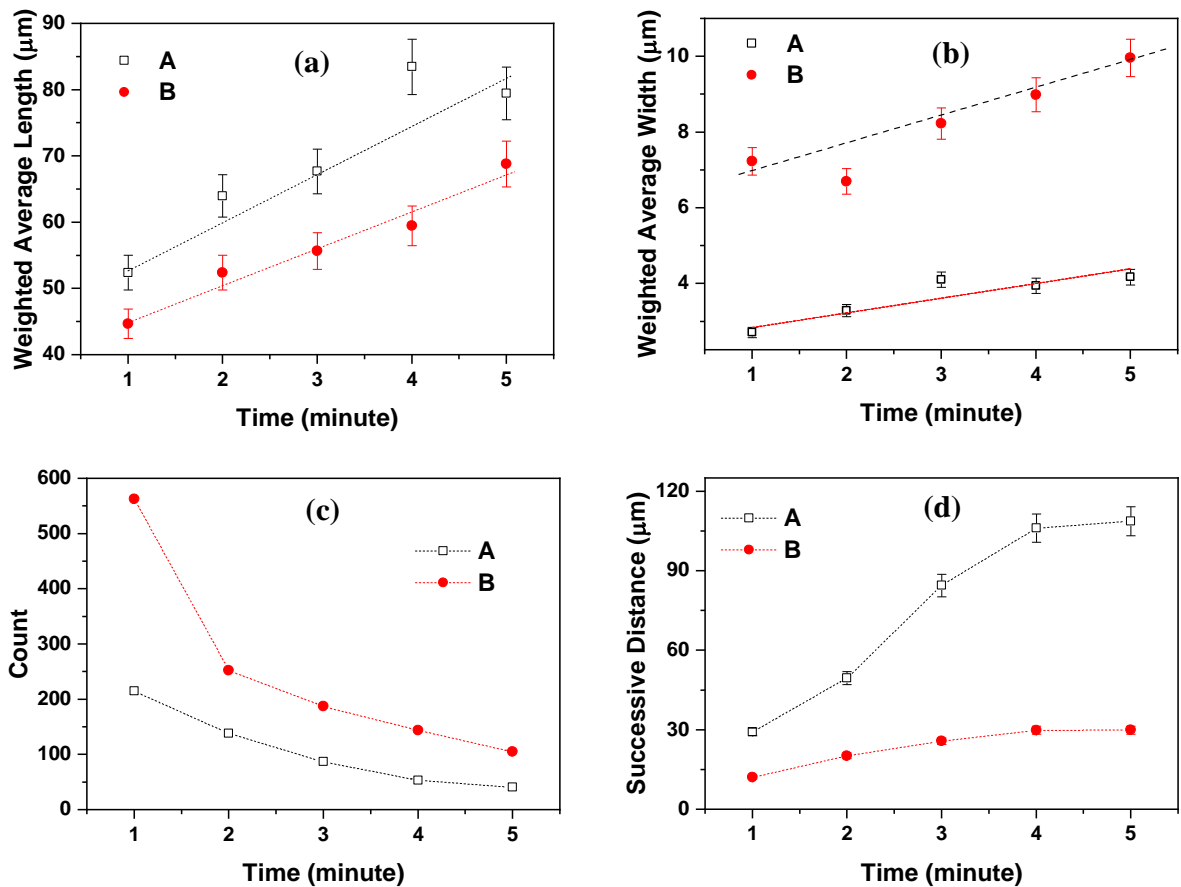

Figure 7: Microstructure parameters - (a) weighted average length, (b) weighted average width, (c) associated counts, and (d) successive distance obtained using ImageJ software as a function of time.

Figure 7(d) shows an increase in the successive distance between the chains with an increase in time. The inter-chain distance increases from 30 µm to 108 µm in the magnetic fluid A, while it varies from 12 µm to 30 µm in the magnetic fluid B. This distribution shows heterogeneity and homogeneity in the samples. It is inferred from Figure 6(a-d) that magnetic fluid A has a heterogeneous distribution of long and thick chains. The structure formed in the fluid B is relatively homogeneous with long and thin chains. It is also concluded that time dependent stability of magnetic fluid B is better under a constant magnetic field compared to the fluid A.

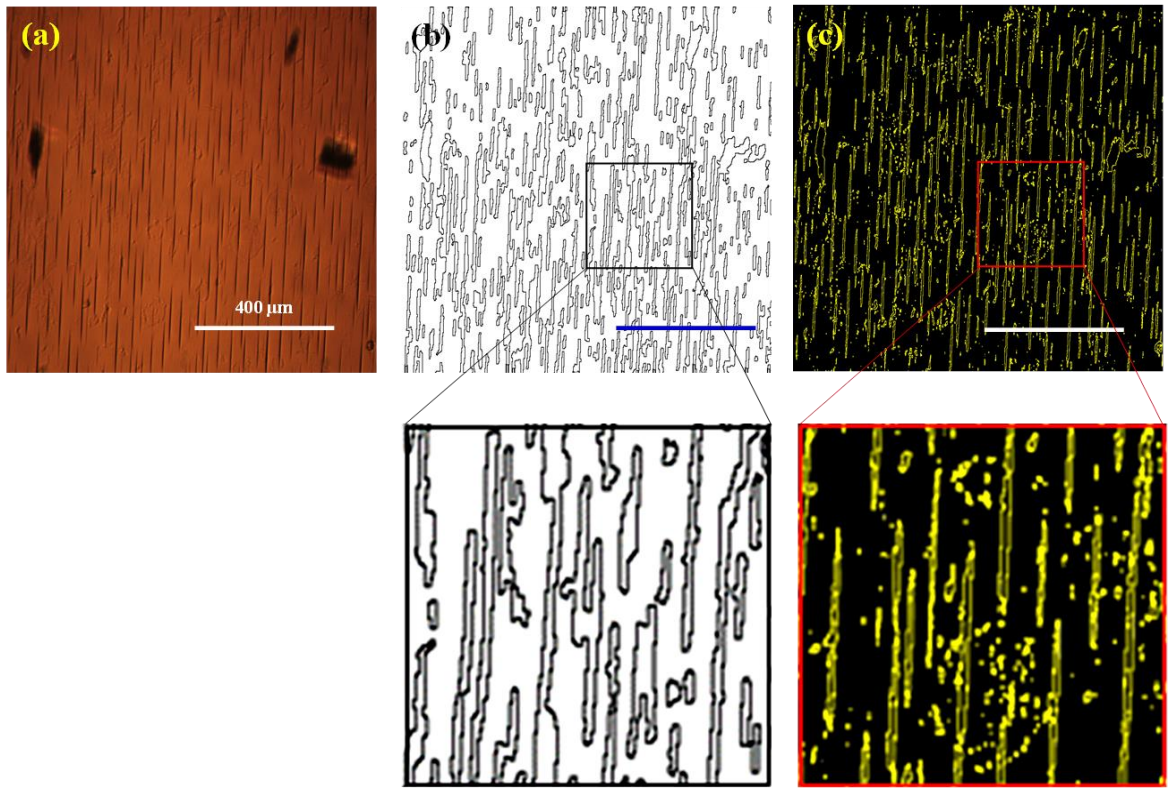

Figure 8: (a) Image captured using a microscope, image process using (b) JColloids method, and (c) proposed method. The scale bar is 400 μm. Adjacent images are the magnified view of the selected area from both the figures (b & c).

The result of the present method is being compared with the JColloids image identification method. Figure 8(a) shows a microscopic image of magnetic fluid B after exposing 0.046 T field for two minutes. This image is analyzed using JColloids method (figure 8b) and the proposed method (figure 8c). It is evident from the figures 8(b & c) that both methods identify the structures, but the structure recognized by both the methods are different. In order to understand the difference clearly, a small section is selected and magnified, and shown adjacently. Prominent structures are observed following both the methods, but the presence of small chains/structures is noticeable using the proposed method. Furthermore, small structures identified using the proposed methods are quite expected, which is not being detected using other methods. JColloid identifies a group of small chains or overlapped chains that are eventually aligning in the magnetic field direction as a single structure. Such large structures are usually anticipated in the magnetorheological fluids but not in the magnetic fluid. The method proposed here is clearly distinguishing between the neighboring chains and identifies as a separate structure. The present method would serve the purpose of determining chain parameters along with chain separation distance; these are essential parameters for developing diffraction grating and other prominent optical devices. The methods reported here can be extended to identify and determine structure formation in the magnetic or nonmagnetic nanofluids.

The magnetic field induced structure parameters are useful to determine diffraction grating angle. The estimated number of lines in a grating per meter ($N = l/d$) (l= length of a grating (say 1 m), and d = interchain distance (in meter)) is in the range of 90-350 and 330-850 respectively for the magnetic fluid sample A and B. Although we need to achieve better control on these deviations, but resultant range of sample A is relatively less. The diffraction angle ($\theta_r$) is calculated as, $\sin \theta_r = m\lambda/d$, where m = order to diffraction, and $\lambda$ = wavelength of incident light (here 650 nm). It infers that in case of first order diffraction (m=±1), $\theta_r$ ranges from 0.69°-2.55°and 2.49° – 6.15° respectively for the Magnetic fluid sample A and B. These preliminary results are encouraging. Further, field dependent experiments will throw some more light to develop tunable diffraction grating.

**CONCLUSION**

The water-based magnetic fluid demonstrates explicit structure formation in the presence of the magnetic field. Despite digital processes, it is difficult to identify and determine structure/chain parameters in the magnetic fluid. ImageJ provides a platform to analyze such structures. Here we proposed MFCPji and MFIDji macros to identify and determine chain parameters, such as chain length, chain width, and associated counts, along with the successive distance of the chains. The chain parameters determined for the aqueous magnetic fluid matches to that of existing literature. Comparing the outcome of the proposed method to that of JColloids method, it is inferred that the presence of small chains can be identify using the proposed method. This chains behaves like diffraction grating lines. The preliminary results of grating parameters are encouraging and has potentiality to develop as tunable diffraction grating. The method can be explored to understand structure formations in any magnetic or nonmagnetic nanofluids. Since automated methods consequently faster, we provide an easy and freely available access to MFCPji and MFIDji for non-commercial use at:

https://github.com/urveshsoni/ImageJ---Macros
https://ruchadesailab.wordpress.com/publication/

ACKNOWLEDGMENT

A special Thanks to Mr. Ashish Gor, Assistant Proffessor, Department of Computer Engineering, Faculty of Technology, Dharmsinh Desai Universiry, Gujarat, India for fruitful discussion related to basic of the Java programming. The work is carried out under grant No: EMR/2016/002278 sanctioned by Science and Engineering Research Board (SERB), Department of Science and Technology (DST), India.